\documentclass[floatfix,twocolumn,aps,prl,amsmath,amssymb]{revtex4}
\usepackage[latin1]{inputenc}
\usepackage[dvips]{graphicx}
\usepackage{epstopdf}
\usepackage{dcolumn}
\usepackage{bm}
\usepackage{color}
\usepackage{url}
\usepackage[colorlinks=true ,citecolor=blue]{hyperref}
\usepackage{amsmath,amssymb}

\begin{document}

\title{Emerging Quantum Hall Effect in Massive Dirac Systems}

\author{Leandro O. Nascimento$^{1,2,3}$, E. C. Marino$^2$, Van S\'ergio Alves$^{1,4}$, and C. Morais Smith$^1$}
\affiliation{$^1$Institute for Theoretical Physics, Centre for Extreme Matter and Emergent Phenomena, Utrecht University, Postbus 80.089, 3508TB Utrecht, The Netherlands \\
$^2$Instituto de F\'\i sica, Universidade Federal do Rio de Janeiro, C.P. 68528, Rio de Janeiro RJ, 21941-972, Brazil \\
$^3$Faculdade de Ci\^encias Naturais, Universidade Federal do Par\'a, C.P. 68800-000, Breves, PA,  Brazil \\
$^4$Faculdade de F\'\i sica, Universidade Federal do Par\'a, Av.~Augusto Correa 01, Bel\'em PA, 66075-110, Brazil. }

\date{\today}

\begin{abstract}
An interaction-driven nonzero quantum Hall conductivity is shown to occur in time-reversal symmetric massive Dirac materials, in the absence of any external agent. The effect is produced through the dynamical breakdown of time-reversal symmetry, which is generated by quantum fluctuations, when the full dynamical electromagnetic interaction among the electrons is taken into account. The manifestation of this emergent parity anomaly should be observed in materials such as silicene, stanene, germanene and transition metal dichalcogenides, at low enough temperatures.

\end{abstract}

\maketitle

{\bf Introduction.} The synthesis of graphene \cite{grapexp} has provided a material which, besides being intrinsically two dimensional (2D), is an example of electronic kinematics governed by the massless Dirac equation. Since then, several compounds have been shown to exhibit a similar honeycomb structure, but with carbon replaced by different atoms. Among these,
silicene \cite{silexp}, germanene \cite{germanene}, and stanene \cite{stanene} stabilize in a buckled structure by virtue of the electrostatic repulsion between the corresponding nuclei, which pushes them to an equilibrium position that alternates between different planes \cite{Silger}. The buckling increases the intrinsic spin-orbit coupling, and although it is still negligible in silicene (of the order of 20 Kelvin \cite{SOCsilicene}), it becomes stronger in germanene and stanene, which have a larger atomic mass. Hence, a full variety of gap sizes is at reach: whereas graphene is gapless, silicene exhibits a gap when grown on substrates or when an external electric field is applied perpendicularly to the plane \cite{Ezawa, Nanotech}, and stanene has a large intrinsic gap, which gives rise to a room temperature quantum spin Hall effect \cite{stanene}.
A second type of similar materials are transition-metal dichalcogenides (MoS$_2$ or WSe$_2$, for example), which are gapped because inversion symmetry is broken by the two different atoms in the unit cell, in addition to the sizable spin-orbit coupling.
Despite their individual peculiarities, a tight-binding approach leads to a two-dimensional massive Dirac description of the low-energy electronic spectrum for all these different materials, which are referred to as massive Dirac systems.

Although the single-particle behavior of these gapped materials has been intensively investigated during the last years, the effect of interactions remained largely unexplored. Due to the mismatch between the 2D dynamics of the electrons and the 3D nature of the photons mediating their interactions, it is necessary to describe these hybrid systems by a projection of QED$_{3+1}$ onto the plane, the so-called Pseudo-QED \cite{marino} (sometimes also referred to as reduced QED \cite{Miransky,Teber1}). This projection leads to a Lagrangean that is non-local in space and time, although the theory respects causality \cite{causal} and remains unitary \cite{unit}.

The investigation of interactions in massless Dirac systems like graphene has been usually based on the use of the static Coulomb interaction, justified by the fact that the Fermi velocity is 300 times less than the speed of light. Although the static description has been able to capture the behavior of the optical conductivity \cite{Ziegler,Vladimir} and of the renormalization of the Fermi velocity $v_F$ \cite{Maria,Elias}, for instance, yet certain phenomena, such as the emergence of anomalies, are out of the reach of purely static theories and require the inclusion of the full dynamical electromagnetic gauge field $A_\mu$.

A well-known example is the axial anomaly related to the chiral U(1) symmetry exhibited by massless Dirac fermions. Classically, the system is invariant under opposite phase transformations performed in the left and right Weyl components of the Dirac fermion. This symmetry is broken by quantum fluctuations, as predicted theoretically \cite{nn, ss, aab} and recently experimentally observed in  Weyl semi-metals, such as TaAs, NbAs, and TaP \cite{weyl1,weyl2}.

Another well-known anomaly is the parity anomaly, which occurs in massless Dirac systems in two-spatial dimensions. Such systems are invariant, at a classical level, under the parity operation. However, when radiative corrections such as the vacuum polarization tensor are evaluated, a parity violating term is invariably generated \cite{Luscher}. The parity anomaly has important consequences for the Hall conductivity because the current-current correlation function, a key ingredient for obtaining the conductivity in the linear-response regime, is given precisely by the vacuum polarization tensor.

In the case of {\it massive} Dirac systems, time-reversal symmetry is usually broken by the mass term. However, due to the existence of two valleys precisely related by such symmetry in these honeycomb structured materials, it is possible to preserve time-reversal  symmetry, provided the bare masses of the two valleys have opposite sign. 

Here, we show that for these time-reversal symmetric massive (gapped) Dirac systems, the parity anomaly does emerge from the dynamical effect of the electromagnetic interaction, which manifests as a different number of counter-propagating current states, associated to each valley. Thus, time-reversal symmetry is broken without the requirement of any external agent that would explicitly break it from the very outset. 
By using the Kubo formula in the framework of Pseudo-QED, we find a quantized Hall conductivity $\sigma_{xy} = 2e^2/h$, characteristic of the integer quantum Hall effect, which however emerges here in the {\it absence} of any applied magnetic field. This is an exact and universal result, grounded on the fact that the two valleys ($K$ and $K$') are related by time-reversal conjugation. 

Such effect should be promptly observed in gapped Dirac materials, like silicene, stanene, or transition metal dichalcogenides with state-of-the-art transport techniques, provided that the sample is clean enough and the temperature low enough. Our findings  are further corroborated by non-perturbative results of the Schwinger-Dyson equation for massive, time-reversal invariant Dirac systems in 2+1D, which show that a parity anomaly occurs, thus generating quantized midgap states \cite{excitons}. These are such that one valley has an extra energy mode of negative energy, whereas the other valley has an extra level with a positive energy, thus dynamically breaking time-reversal invariance.

{\bf The Model.}
Let us start by considering an electronic system on a honeycomb lattice, where the symmetry between the $A$ and $B$ sublattices is broken either by a different chemical potential or by the presence of different atoms in each of them. Within a tight-binding description, the spectrum shows a massive Dirac dispersion around each of the $K$ and $K'$ vertices of the Brillouin zone, herewith called valleys. Then, we assume that the Dirac electrons interact through the full dynamical electromagnetic interaction, which in 2+1D is described by  PQED \cite{marino}.
The Lagrangian model describing the system is given by
\begin{equation}
{\cal L}= \frac{1}{2} F_{\mu \nu}\left(\frac{1}{\sqrt{-\Box}}\right) F^{\mu\nu}+\bar\psi_a (i\partial\!\!\!/-M_{a}) \psi_a+j^{\mu}A_{\mu},
\label{action}
\end{equation}
where $i\partial\!\!\!/ =i\gamma^0\partial_0+i\,v_F\gamma^i\partial_i$ and $j^\mu=e\,\bar\psi\gamma^\mu\psi=e\,(\bar\psi\gamma^0\psi,v_F \,\bar\psi\gamma^i\psi)$.
Here, $\psi_a=(\psi_A,\psi_B)_a$ is a two-component Dirac field, corresponding to the inequivalent $A$ and $B$ sublattice sites of the honeycomb lattice, $a=K\uparrow, K\downarrow, K'\uparrow, K'\downarrow$ is a flavor index accounting for the spin and valley internal degrees of freedom, $\bar\psi_a =\psi_a^\dagger\gamma^0$, with $\gamma^\mu$ rank-2 Dirac matrices, $A_\mu$ is the U(1) gauge field, which intermediates the electromagnetic interaction in 2D (pseudo electromagnetic field), and $F_{\mu \nu}$ is the usual field-intensity tensor corresponding to it. The coupling constant $e^2= 4\pi\alpha$ is conveniently written in terms of $\alpha$, the fine-structure constant in natural units.

Complying with previous studies \cite{semenoff,Ezawa,wsil}, the mass term is written as $M_{a}=\xi \Delta$, and without loss of generality, we choose the valley index $\xi$ to be positive ($\xi=+1$) for valley $K$ and negative ($\xi=-1$) for valley $K'$.
Hence, $M_{a}=(M_{K\uparrow},M_{K\downarrow},M_{K'\uparrow},M_{K'\downarrow})= (\Delta,\Delta,-\Delta,-\Delta) $ and we consider, e.g., $\Delta<0$.
 
This choice for the masses lets the Lagrangean (\ref{action}) invariant under time-reversal symmetry. Indeed, under this symmetry operation $\bar\psi_\xi  \psi_\xi \rightarrow - \bar\psi_{-\xi}  \psi_{-\xi }$, and consequently 
\begin{equation}
\sum_{a} M_{a} \bar\psi_a \psi_a= \sum_{s=\uparrow,\downarrow}
\sum_{\xi=\pm 1}\xi \Delta  \bar\psi_{s,\xi} \psi_{s\xi}
\end{equation}
is clearly time-reversal invariant. From Eq.~(\ref{action}), we have that the energy dispersion of the Dirac particles is $E(\textbf{p})=\pm  \sqrt{v_F^2 \textbf{p}^2+\Delta^2}$, which reproduces the tight-binding theory for silicene \cite{semenoff,Ezawa}, for example, and corresponds to a gap $2 \Delta$.

{\bf Kubo formula.}
Our aim is to investigate the transport properties of the system described above. We, therefore, use the Kubo formula in order to calculate the $\omega \to 0$ limit of the optical conductivity (dc-conductivity), for each flavor
\begin{equation}
\sigma_{a}^{ik}= \lim_{\omega\rightarrow 0, \textbf{p}\rightarrow 0}\frac{i \langle j^i j^k \rangle _{a}}{\omega}=\lim_{\omega\rightarrow 0, \textbf{p}\rightarrow 0}\frac{\Pi_{a}^{ik}(\omega,v_F \textbf{p})}{\omega}.
\label{kubosup}
\end{equation}
The vacuum polarization $\Pi^{\mu\nu}$ has been calculated by Coste and Luscher
for the case of a single two-component Dirac fermion. The Euclidean one-loop result for
an electron with flavor $a$, in units of $e^2 / \hbar$,
reads \cite{Luscher}
\begin{eqnarray}
i\Pi_{a}^{ij} (\omega, v_F\textbf{p}) &=& A_a(p^2)\left [\delta^{ij}p^2 -\textbf{p}^i  \textbf{p}^j\right]+
\nonumber \\
&\ &
B_a(p^2) \epsilon^{ij} \omega,
\label{1336} 
\end{eqnarray}
in which $p^2=\omega^2-v_F^2|\textbf{p}|^2$ and
\begin{eqnarray}
A_a(p^2) &=& \frac{1}{2\pi} \int_0^1 dx \, \frac{x(1-x) p^2}{\sqrt{M_a^2+x(1-x)p^2}},
\nonumber \\ 
B_a(p^2) &=&  \frac{\xi}{2\pi}\left[n+\frac{1}{2}\int_0^1 dx \left(1-\frac{M_a}{\sqrt{M_a^2+x(1-x)p^2}}\right)\right], 
\nonumber \\
\label{1337}
\end{eqnarray}
where the $\xi=\pm 1$ sign corresponds, respectively, to the contributions of valleys $K$ and $K'$.

Now, considering the limits, we find
\begin{eqnarray}
A_{a}(p^2) &\stackrel{\omega; |\textbf{p}|\rightarrow 0}\longrightarrow & C\omega^2 \rightarrow 0
\nonumber \\ 
B_{a}(p) &\stackrel{\omega; |\textbf{p}|\rightarrow 0}\longrightarrow&\frac{\xi}{2\pi}\left[n+\frac{1-\mathrm{sgn}(M_a)}{2}\right], 
\label{1338a}
\end{eqnarray}
where $C$ is a constant and $\mathrm{sgn}(M_a)$ is the sign function.

We may define two conductivities: the total conductivity and the valley conductivity, respectively, given by
\begin{equation}
\sigma^{ik}_{\rm{T}}=\sum_a \sigma^{ik}_a, \ \ \ \ \ \ \sigma^{ik}_{\rm{V}}=\sum_a \xi_a \sigma^{ik}_a=
 \sum_{s=\uparrow,\downarrow}\left[\sigma^{ik}_{s,K}- \sigma^{ik}_{s,K'} \right].
\end{equation}

Using Eqs.~(\ref{1338a}), we obtain the longitudinal and transverse components of the
conductivity for each flavor, in the case $\Delta \neq 0$:
\begin{eqnarray}
\sigma^{xx}= 0 \quad a=K\uparrow, K\downarrow, K'\uparrow, K'\downarrow
\label{long}
\end{eqnarray}
and
\begin{eqnarray}
\sigma^{xy}=\left\{\begin{array}{rc}
+ (n+1) e^2/h, &  \quad a=K\uparrow, K\downarrow\\
-n e^2/h, & \quad a = K'\uparrow, K'\downarrow \\
\end{array}\right. \label{transA}
\end{eqnarray}
respectively.
Note that all $v_F$-dependence on the Kubo formula disappears when we perform the limit $\textbf{p}\rightarrow 0$.

By substituting these results in the previous expressions for the total and valley conductivities and summing over all flavors, we find
\begin{eqnarray}
\sigma_{\mathrm{T}}^{xy}=2e^2/h\ \ \ 
; \ \ \  \sigma_\mathrm{V}^{xy}=4(n+1/2)e^2/h.
\label{trans22}
\end{eqnarray}

This is the main result of this Letter: the parity anomaly, which is the transverse term in the vacuum polarization tensor, leads to a finite Hall conductivity, even in the absence of any applied magnetic field. This occurs in a system that was, from the beginning, time-reversal invariant, and the symmetry has been broken dynamically. Notice that this result is exact because according to the Coleman-Hill theorem \cite{ch}, there are no corrections arising from higher-order loops to the transverse term in the vacuum polarization tensor.

Transverse conductivity in a topological insulator is related to the number of conducting edge states and, through the bulk-boundary relation, to the number of gapped states in the bulk. Hence, we immediately infer, from (\ref{transA}), that the time-reversal symmetry is dynamically broken, since the number of propagating states from valley $K$ differs from that of valley $K'$. Since each of these propagates in opposite directions, one can predict the emergence of a net edge current.

For completeness, we include the zero-gap case $(\Delta = 0)$, which would befit graphene:
\begin{eqnarray}
\sigma_0^{xx}=\left\{\begin{array}{rc}
\pi e^2/8h, & \Delta =0, \quad a=K\uparrow, K\downarrow\\
\pi e^2/8h, &\Delta =0, \quad a=K'\uparrow, K'\downarrow
\end{array}\right. \label{long1}
\end{eqnarray}
and
\begin{eqnarray}
\sigma_0^{xy}=\left\{\begin{array}{rc}
 -(n+1/2)e^2/h,&\Delta=0, \quad a= K\uparrow, K\downarrow \\
 (n+1/2)e^2/h,& \Delta=0, \quad a= K'\uparrow, K'\downarrow,
\end{array}\right. \label{trans}
\end{eqnarray}
which yield, after summing upon the flavors \cite{prx}
\begin{eqnarray}
\sigma_{0\mathrm{T}}^{xy}&=&0, \ \ \ \ \ \ \ \ 
 \ \ \ \ \ \ \ \ \   \sigma_{0\mathrm{V}}^{xy}=4(n+1/2)e^2/h,
\nonumber \\
\sigma_{0\mathrm{T}}^{xx}&=&\frac{\pi}{2}e^2/h, \ \ \ \ \ \ \ \ \ \ \ \ \ \ \ 
 \sigma_{0\mathrm{V}}^{xx}=0.
\label{trans22x}
\end{eqnarray}
Now both valleys, $K$ and $K'$ yield the same number of states and time-reversal symmetry is preserved. The total transverse conductivity vanishes, even though a nonzero valley transverse conductivity is predicted \cite{prx}.

\begin{figure}[htb]
\includegraphics[scale=1.0]{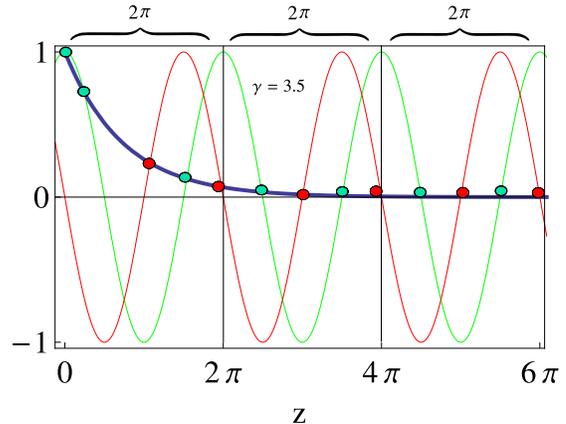}
\caption{Graphical representation of solutions $z_{\pm}$ of the transcendental Eq.~(\ref{2}). The blue line is the exponential function in the lhs of Eq.~(\ref{2}) for $\gamma=3.5$ (an artificial value chosen to improve the visualization). The green line depicts $+\cos z_{+}$ and the red one $-\sin z_{-}$. The roots $\{z_+,z_-\}$ are represented by small circles in green and red, respectively. } \label{fig1}
\end{figure}

{\bf Schwinger-Dyson.} Our results for the total and valley transverse conductivities are corroborated by non-perturbative calculations using the Schwinger-Dyson equation in the framework of Pseudo-QED. Indeed, this revealed the existence of a series of quantized midgap energy levels, related to the quantized transverse conductivities \cite{excitons}, in the same way as the Landau levels are related to the transverse conductivity plateaus, in the case of an applied magnetic field.

It has been shown recently that the Schwinger-Dyson equation leads to a differential equation for the self-energy $\Sigma$, the solution of which
allows us to infer the existence of two-series of energy eigenstates, corresponding, respectively, to the eigenvalues $E_n^+= \sqrt{|\textbf{p}|^2 + (M_n^+)^2}$  and  $E_n^-= - \sqrt{|\textbf{p}|^2 + (M_n^-)^2}$ \cite{excitons}. The energy of the bound states is given, respectively, by
\begin{eqnarray}
M_n^+(K')= \Lambda e^{-X_n/\gamma}\ \ \ ;\ \ \  M_n^-(K')=- \Lambda e^{-Y_n/\gamma},
\label{1}
\end{eqnarray}
and by
\begin{eqnarray}
M_n^+(K)= \Lambda e^{-Y_n/\gamma}  \ \ \ ;\ \ \ M_n^-(K)= - \Lambda e^{-X_n/\gamma}.
\label{Mpm} 
\end{eqnarray}
where $X_n$ and $Y_n$ are, respectively, the solutions of the transcendental equations
\begin{eqnarray}
 e^{-3z/2\gamma}=\cos z\ \ \ ;\ \ \  e^{-3z/2\gamma}=-\sin z. 
\label{2}
\end{eqnarray}

In the above expressions,
$2\gamma=\sqrt{\alpha/\alpha_c-1}$, where the fine-structure constant $\alpha$ is supposed to be larger than a critical value $\alpha_c=\pi/8\approx 0.40$.
$\Lambda$ is a natural energy cutoff provided by the finite lattice spacing. A convenient choice would be $\Lambda =  |\Delta|$.

{\bf Cantor method.} 
Next, we shall prove that the total number of solutions for $X_n$ differs from that for $Y_n$. It is clear that these numbers do not depend on $\alpha$, provided that $\alpha>\alpha_c$. 
 In order to count the number of solutions, we use the method of Cantor for counting infinite sets \cite{Cantor}. For this purpose, we make a partition of the total range of $z$ into the following sets:
\begin{eqnarray}
\bigcup_{n=0}^\infty \{(2n)\pi,(2n+2)\pi\}.
\label{4}
\end{eqnarray}

It is not difficult to see that each of the intervals above possesses two  $X_n$ and two $Y_n$ solutions, for $n\geq 1$, while the first interval in (\ref{4}), $n=0$, possesses three $X_n$ and two $Y_n$ solutions, see Fig.1.
 Consequently, we find by induction that the number of solutions for positive mass, $M_n^+$ exceeds the number of solutions for negative mass, $M_n^-$ by one for the valley $K'$, whereas for the valley $K$ is precisely the opposite (see Fig.2 for a sketch of the energy levels). 

This observation shows that there exists a one-to-one correspondence between the midgap states $M_n^+(K)$ and $M_n^-(K')$, as well as between $M_n^+(K')$ and $M_n^-(K)$, with the conductivity quanta associated to each valley $K$ or $K'$.

 For the undoped material, when the chemical potential is at the center of the gap, where $E = 0$, there will be one extra state for each spin orientation in valley $K$.
Since the electrons in the $K$ and $K'$ valleys counter-propagate, the contribution of all filled energy levels vanish, except for the additional level in $K$. The obtained quantized Hall conductivity $\sigma^{xy} = 2 e^2 / h$ is hence a mere consequence of the parity anomaly, combined with the B\"uttiker-Landauer rule that each mode carries one quantum of conductance $\sigma_0 = e^2 /h$ (here multiplied by two because of the spin).

\begin{figure}[tbh]
\centering
\includegraphics[scale=0.15]{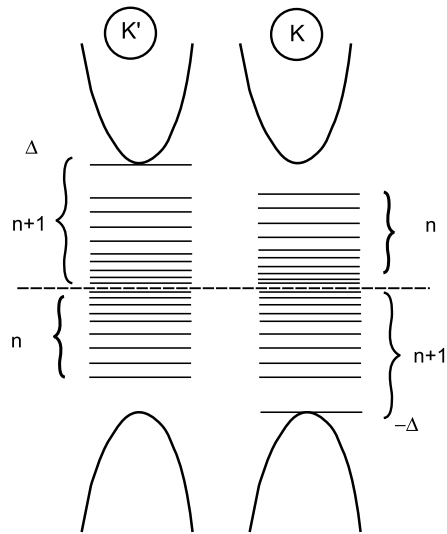}
\caption{Sketch of the number of midgap states in the valleys $K$ and $K'$. The additional unpaired state at the top of the valence band in valley $K$ is responsible for the finite value of the quantum Hall conductivity when the chemical potential is at zero energy. } \label{fig2}
\end{figure}

{\bf Activation Temperature.} 
Now, we determine the activation temperature $T^*$, above which the discreteness of the generated energy levels is destroyed by thermal activation.
According to the Arrhenius law, the activation temperature $T^*$ is given by \cite{arr}
\begin{equation}
T^*=\frac{-\Delta E}{k_B \ln(N/N_0)}, \label{t01}
\end{equation}
where $N/N_0$ expresses the ratio of successful activation events, i.e, the fraction of particles that have been thermally excited from the ground state $E_0$ to the first-excited state with energy $E$, with $\Delta E = E - E_0$.

The splitting between the first two lowest energy levels provides a reasonable estimate for it. Typical values for transition metal dichalcogenides are of the order of $\Delta E = E - E_0 \sim 0.10$ eV. The usual value for the ratio of succesful events is $N/N_0\approx 10^{-14}$ \cite{arr}. Note, however, that $T^*$ is weakly dependent on this ratio because of the logarithm. Indeed, $\ln N/N_0\approx -30$, hence
\begin{equation}
k_B T^*\approx \frac{0.10}{ 30}\approx 0.0033. 
\label{t02}
\end{equation}
This corresponds to an activation temperature of the order $T^*\approx 30 $K. Broadening of the energy levels by disorder will further reduce this value. 

{\bf Conclusions.}
The symmetries exhibited by the Hamiltonian of a physical system are not always revealed by experimental observations. Indeed, it may happen that some of them are destroyed by the existence of asymmetric vacuum states, which would spoil any associated conservation law that might exist. 

It may also happen that a symmetry the system may exhibit at a classical level is just spoiled by quantum fluctuations.
The corresponding classical symmetries are, then, said to present an ``anomaly''. Examples of this mechanism are the axial and parity anomalies, which are known to occur in systems containing Dirac excitations.

Here, we predict that a parity anomaly should occur in recently synthesized materials containing massive Dirac excitations, such as silicene, germanene, stanene, and transition metal dichalcogenides. A crucial condition for generating the parity anomaly by radiative corrections is the presence of the full dynamical coupling of the matter current with the electromagnetic gauge field $A_\mu$, namely the full tri-linear QED vertex. In contrast, many recent studies of condensed-matter systems such as graphene, for instance, which are based on a static Coulomb interaction, albeit justified by the fact that $v_F \ll c$, actually miss the effects of the parity anomaly because the full interaction vertex is just not there. Moreover, the static Coulomb interaction artificially breaks the particle-hole symmetry, and as a consequence one solution
is missed when solving the Schwinger-Dyson equation. 

The most important effect of the dynamically generated parity anomaly is the emergence of a finite quantum Hall conductivity even in the absence of an external magnetic field or any other perturbation that would explicitly break time-reversal symmetry whatsoever. In a seminal paper, some years ago, Haldane has shown that the quantum Hall effect actually does not require the presence of a net magnetic field: it would suffice that the time-reversal-symmetry is explicitly broken \cite{Haldane}. Our result, however, goes even further by showing that the quantum Hall effect can be driven by a {\it dynamical} breakdown of time-reversal symmetry, without any previous explicit breaking thereof. 

The effect manifests itself, both by an asymmetry in the contribution to the transverse conductivity originating from each valley and by the appearance of a different number of midgap solutions with positive and negative signs, corresponding respectively to each of the valleys, $K$ and $K'$. Since the currents associated to each valley counter-propagate at the edges, the effect generates a net edge current.
A quantized Hall conductivity, spontaneously induced by such an intrinsic mechanism, which derives from the full dynamical electromagnetic interaction among the electrons, should have profound consequences both in principles and applications.

\acknowledgments
This work was supported in part by CNPq (Brazil), CAPES (Brazil), FAPERJ (Brazil), and by the Brazilian government project Science Without Borders. We are grateful to G. van Miert , M. O. Goerbig, A. Quelle, and N. Menezes for fruitful discussions.

\end{document}